\documentclass[runningheads]{llncs}

\usepackage{graphicx}
\usepackage{ctable}
\usepackage[caption=false]{subfig}

\newcommand{\para}[1]{\vspace{2mm}\noindent\textbf{#1}}

\begin{document}
\title{The Unfairness of Popularity Bias in Music Recommendation: A Reproducibility Study}
\titlerunning{The Unfairness of Popularity Bias in Music Recommendation}
\author{Dominik Kowald \inst{1} \and Markus Schedl \inst{2} \and Elisabeth Lex \inst{3}}
\authorrunning{Dominik Kowald, Markus Schedl and Elisabeth Lex}

\institute{Know-Center GmbH, Graz, Austria \\\email{dkowald@know-center.at}\and
Johannes Kepler University Linz, Austria
\\\email{markus.schedl@jku.at}\and
Graz University of Technology, Graz, Austria  \\\email{elisabeth.lex@tugraz.at}}

\maketitle  

\begin{abstract}
Research has shown that recommender systems are typically biased towards popular items, which leads to less popular items being underrepresented in recommendations. The recent work of Abdollahpouri et al. in the context of movie recommendations has shown that this popularity bias leads to unfair treatment of both long-tail items as well as users with little interest in popular items. In this paper, we reproduce the analyses of Abdollahpouri et al. in the context of music recommendation. Specifically, we investigate three user groups from the Last.fm music platform that are categorized based on how much their listening preferences deviate from the most popular music among all Last.fm users in the dataset: (i) low-mainstream users, (ii) medium-mainstream users, and (iii) high-mainstream users. In line with Abdollahpouri et al., we find that state-of-the-art recommendation algorithms favor popular items also in the music domain. However, their proposed Group Average Popularity metric yields different results for Last.fm than for the movie domain, presumably due to the larger number of available items (i.e., music artists) in the Last.fm dataset we use. Finally, we compare the accuracy results of the recommendation algorithms for the three user groups and find that the low-mainstreaminess group significantly receives the worst recommendations.

\end{abstract}

\section{Introduction}
Recommender systems are quintessential tools to support users in finding relevant information in large information spaces~\cite{ricci2011introduction}. However, one limitation of typical recommender systems is the so-called popularity bias, which leads to the underrepresentation of less popular (i.e., long-tail) items in the recommendation lists~\cite{abdollahpouri2017controlling,brynjolfsson2006niches,jannach2015recommenders}. The recent work of Abdollahpouri et al.~\cite{abdollahpouri2019unfairness} has investigated this popularity bias from the user perspective in the movie domain. The authors have shown that state-of-the-art recommendation algorithms tend to underserve users, who like unpopular items.

In this paper, we reproduce this study and conduct it in the music domain. As described in~\cite{schedl2018current}, there are several aspects of music recommendations that make them different to, e.g., movie recommendations such as the vast amount of available items. Therefore, we investigate music recommendations concerning popularity bias and, for reasons of comparability, raise the same two research questions as in~\cite{abdollahpouri2019unfairness}:
\begin{itemize}
    \item \textit{RQ1}: To what extent are users or groups of users interested in popular music artists?
    \item \textit{RQ2}: To what extent does the popularity bias of recommendation algorithms affect users with different inclination to mainstream music?
\end{itemize}

For our experiments, we use a publicly available Last.fm dataset and address \textit{RQ1} in Section~\ref{sec:rq1} by analyzing the popularity of music artists in the user profiles. Next, we address \textit{RQ2} in Section~\ref{sec:rq2} by comparing six state-of-the-art music recommendation algorithms concerning their popularity bias propagation.

\section{Popularity Bias in Music Data} \label{sec:rq1}
For our reproducibility study, we use the freely available LFM-1b dataset~\cite{schedl2016lfm}. Since this dataset contains 1.1 billion listening events of more than 120,000 Last.fm users and thus is much larger than the MovieLens dataset used in~\cite{abdollahpouri2019unfairness}, we focus on a subset of it. Precisely, we extract 3,000 users that reflect the three user groups investigated in~\cite{abdollahpouri2019unfairness}. To this end, we use the mainstreaminess score, which is available for the users in the LFM-1b dataset and which is defined as the overlap between a user's listening history and the aggregated listening history of all Last.fm users in the dataset~\cite{bauer2019global}. It thus represents a proxy for a user's inclination to popular music.

\begin{figure}[t]
\centering
   \subfloat[Long-tail of listening counts.]{
      \includegraphics[width=.44\textwidth]{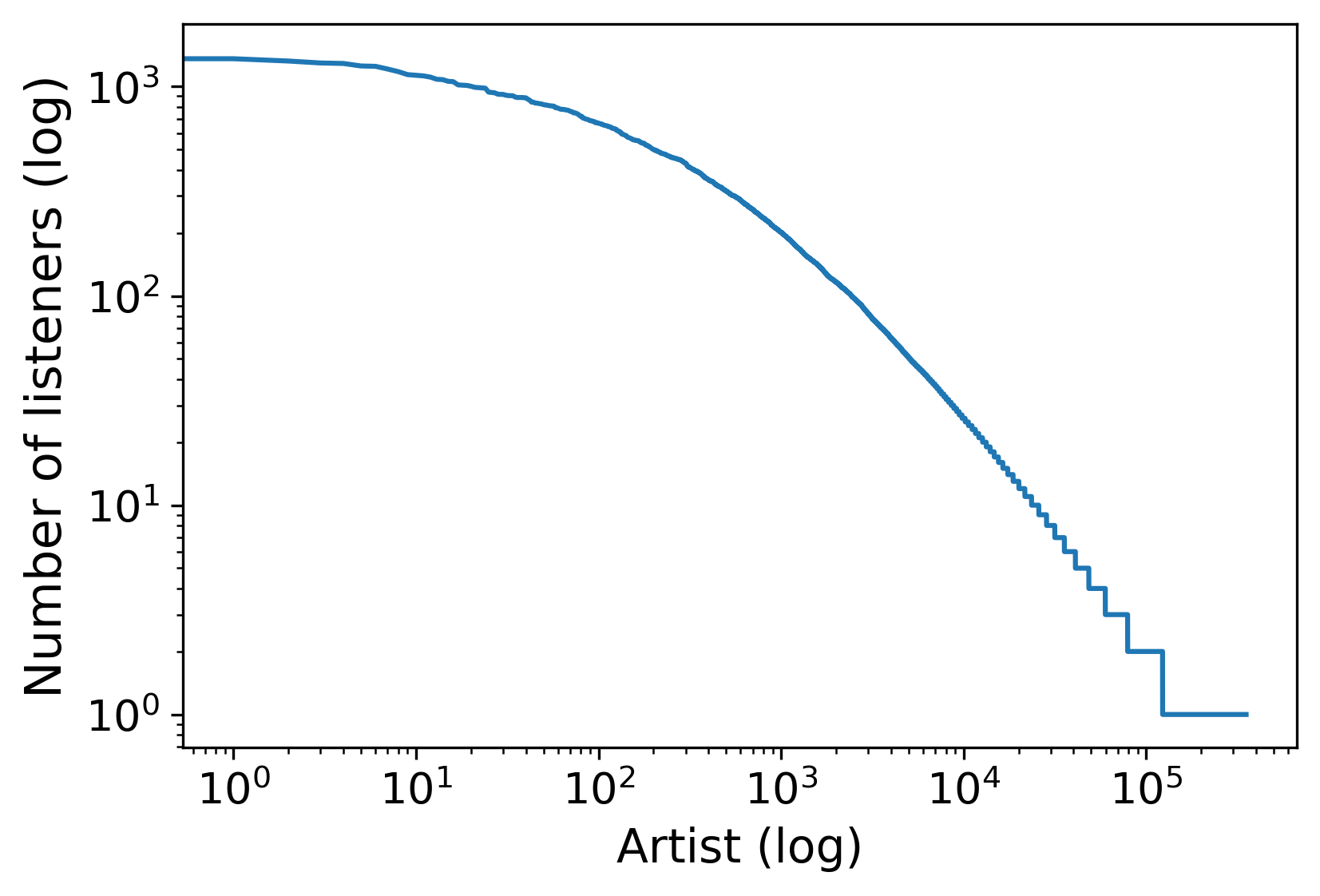} \label{fig:dataset-a}} 
~
   \subfloat[Popular artists in user profiles.]{
      \includegraphics[width=.44\textwidth]{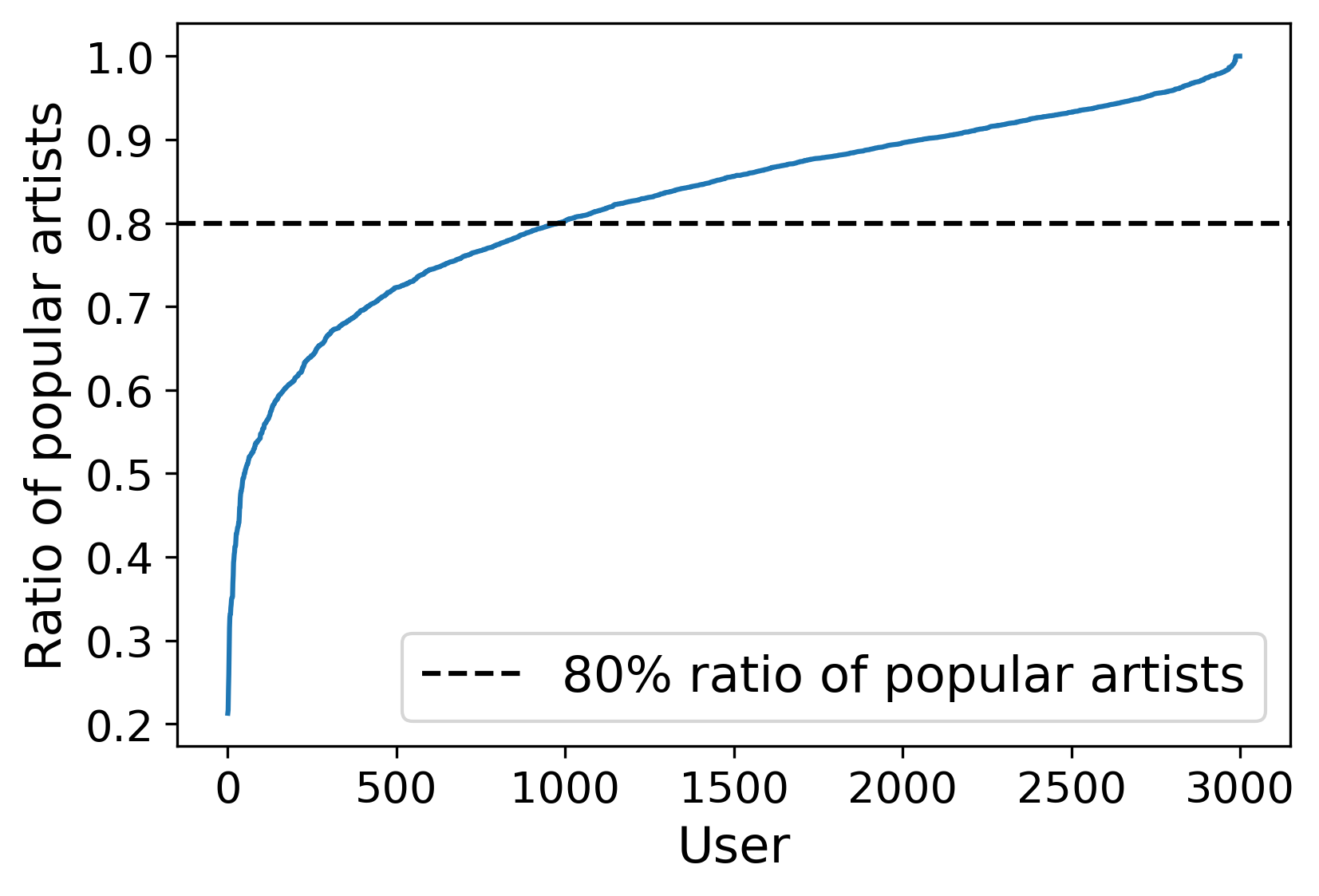} \label{fig:dataset-b}} 
   \caption{Listening distribution of music artists. We find that around 1/3 (i.e., 1,000) of our users actually listen to at least 20\% of unpopular artists.
   \vspace{-3mm}}
   \label{fig:dataset}
\end{figure}

Our subset consists of the 1,000 users with lowest mainstreaminess scores (i.e., the LowMS group), the 1,000 users with a mainstreaminess score around the median (i.e., the MedMS group), and the 1,000 users with the highest mainstreaminess scores (i.e., the HighMS group). In total, we investigate 1,755,361 user-artist interactions between 3,000 users and 352,805 music artists. Compared to the MovieLens dataset with only 3,900 movies that Abdollahpouri et al.~\cite{abdollahpouri2019unfairness} have used in their study, our itemset is, consequently, much larger.

\para{Listening distribution of music artists.}
Figure~\ref{fig:dataset} depicts the listening distribution of music artists in our Last.fm dataset. As expected, in Figure~\ref{fig:dataset-a}, we observe a long-tail distribution of the listener counts of our items (i.e., artists). That is, only a few artists are listened to by many users, while most artists (i.e., the long-tail) are only listened to by a few users. Furthermore, in Figure~\ref{fig:dataset-b}, we plot the ratio of popular artists in the profiles of our 3,000 Last.fm users. As in~\cite{abdollahpouri2019unfairness}, we define an artist as popular if the artist falls within the top 20\% of artists with the highest number of listeners.
We see that around 1,000 of our 3,000 users (i.e., around 1/3) have at least 20\% of unpopular artists in their user profiles. 
This number also corresponds to the number of low-mainstream users we have in the LowMS user group.

\begin{figure}[t]%[ht]
\centering
   \subfloat[Number of popular artists.]{
      \includegraphics[width=.44\textwidth]{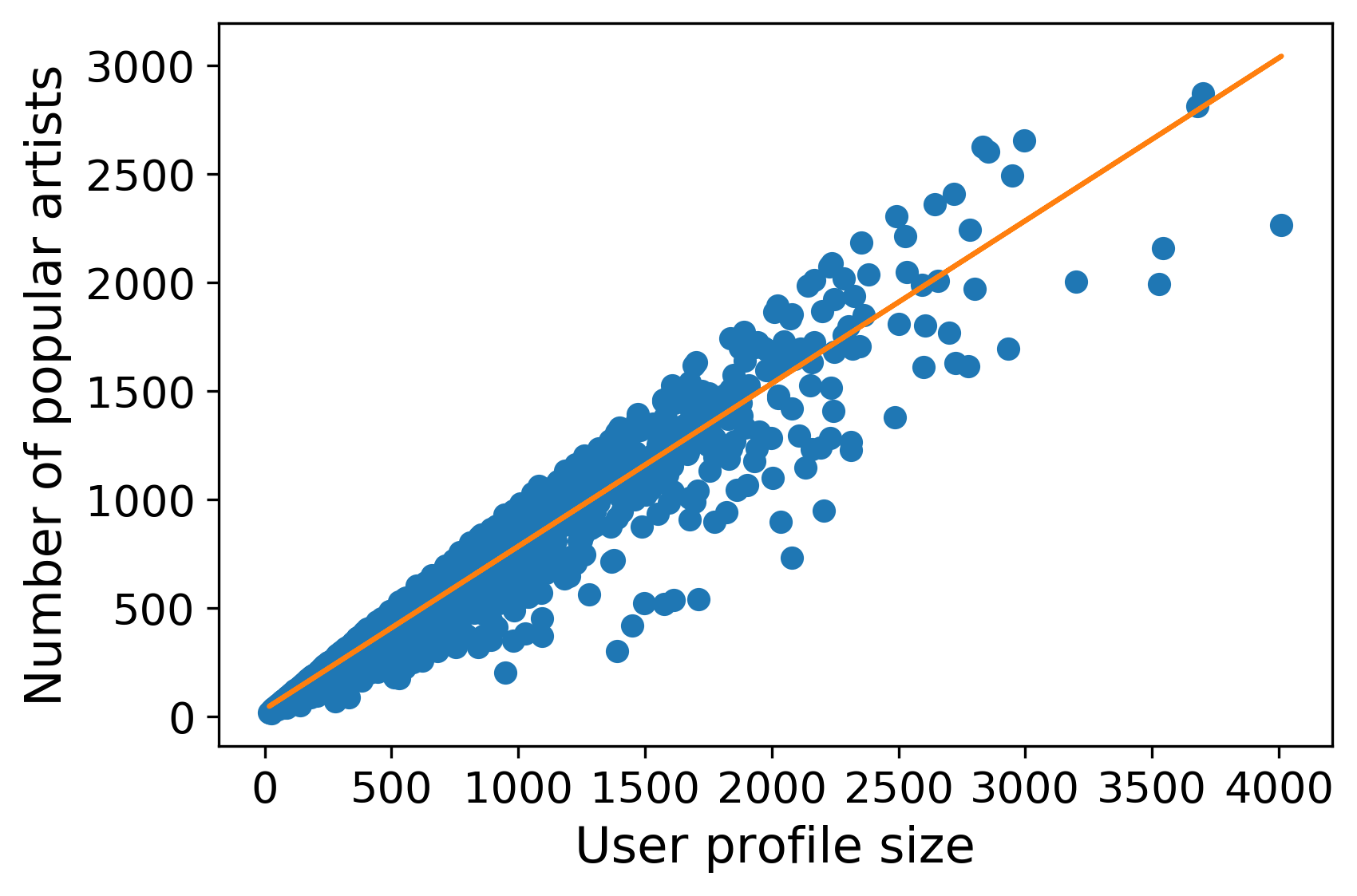} \label{fig:correlations-a}}
~
   \subfloat[Average popularity of artists.]{
      \includegraphics[width=.44\textwidth]{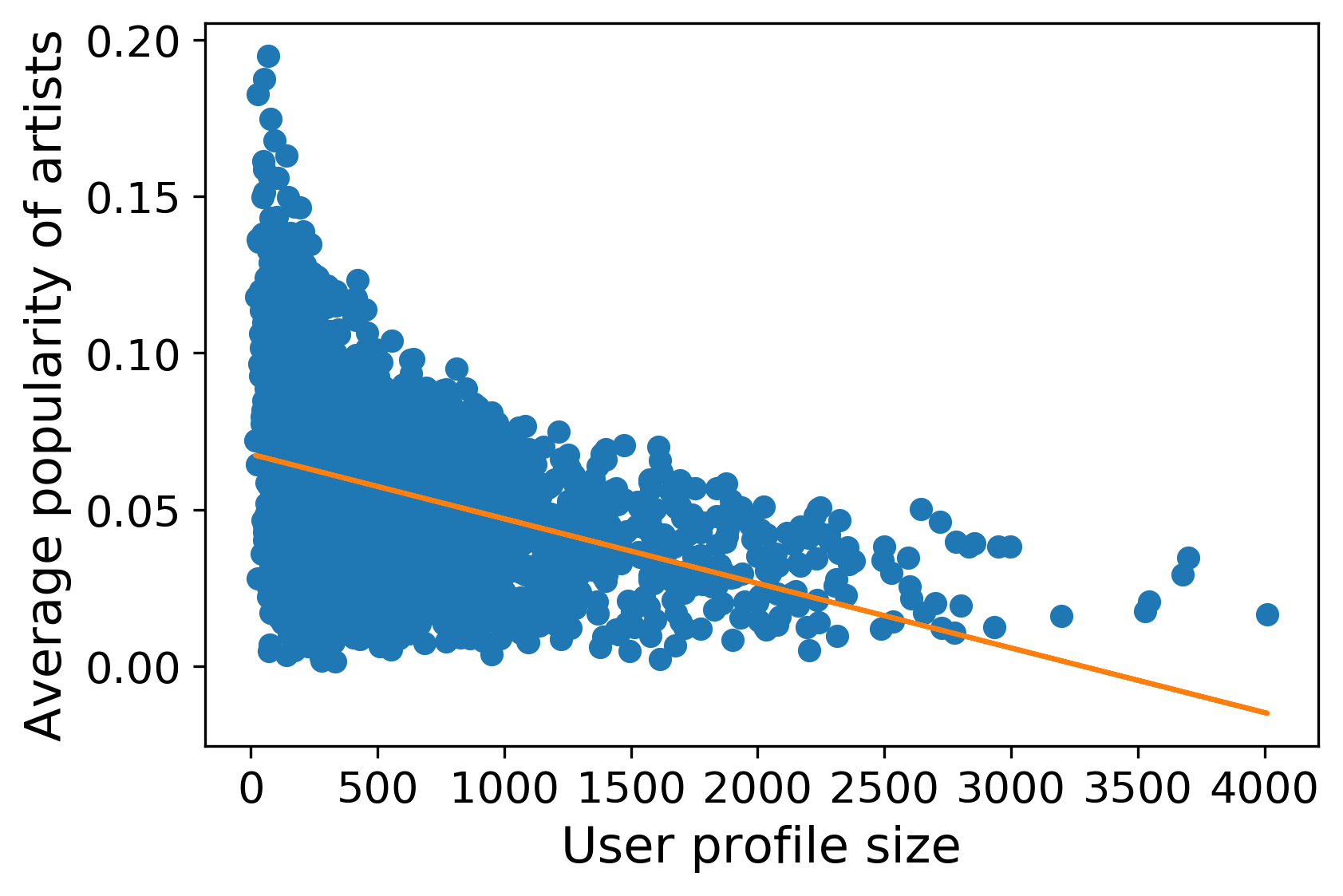} \label{fig:correlations-b}}
   \caption{Correlation of user profile size and the popularity of artists in the user profile. While there is a positive correlation between profile size and number of popular artists, there is a negative correlation between profile size and the average artist popularity.
   \vspace{-3mm}}
   \label{fig:correlations}
\end{figure}

\para{User profile size and popularity bias in music data.}
Next, in Figure~\ref{fig:correlations}, we investigate if there is a correlation between the user profile size (i.e., number of distinct items/artists) and the popularity of artists in the user profile. Therefore, in Figure~\ref{fig:correlations-a}, we plot the number of popular artists in the user profile over the profile size
As expected, we find a positive correlation ($R = .965$) since the likelihood of having popular artists in the profile increases with the number of items in the profile. However, when plotting the average popularity of artists in the user profile over the profile size in Figure~\ref{fig:correlations-b}, we find a negative correlation ($R = -.372$), which means that users with a smaller profile size tend to listen to more popular artists. 
As in~\cite{abdollahpouri2019unfairness}, we define the popularity of an artist as the ratio of users who have listened to this artist.

Concerning \textit{RQ1}, we find that one-third of our Last.fm users have at least 20\% of unpopular artists in their profiles and thus, are also interested in low-mainstream music. Furthermore, we find that users with a small profile size tend to have more popular artists in their profiles than users with a more extensive profile size. These findings are in line with what Abdollahpouri et al. have found~\cite{abdollahpouri2019unfairness}.

\begin{figure}[t]
\centering
   \subfloat[Random.]{
      \includegraphics[width=.32\textwidth]{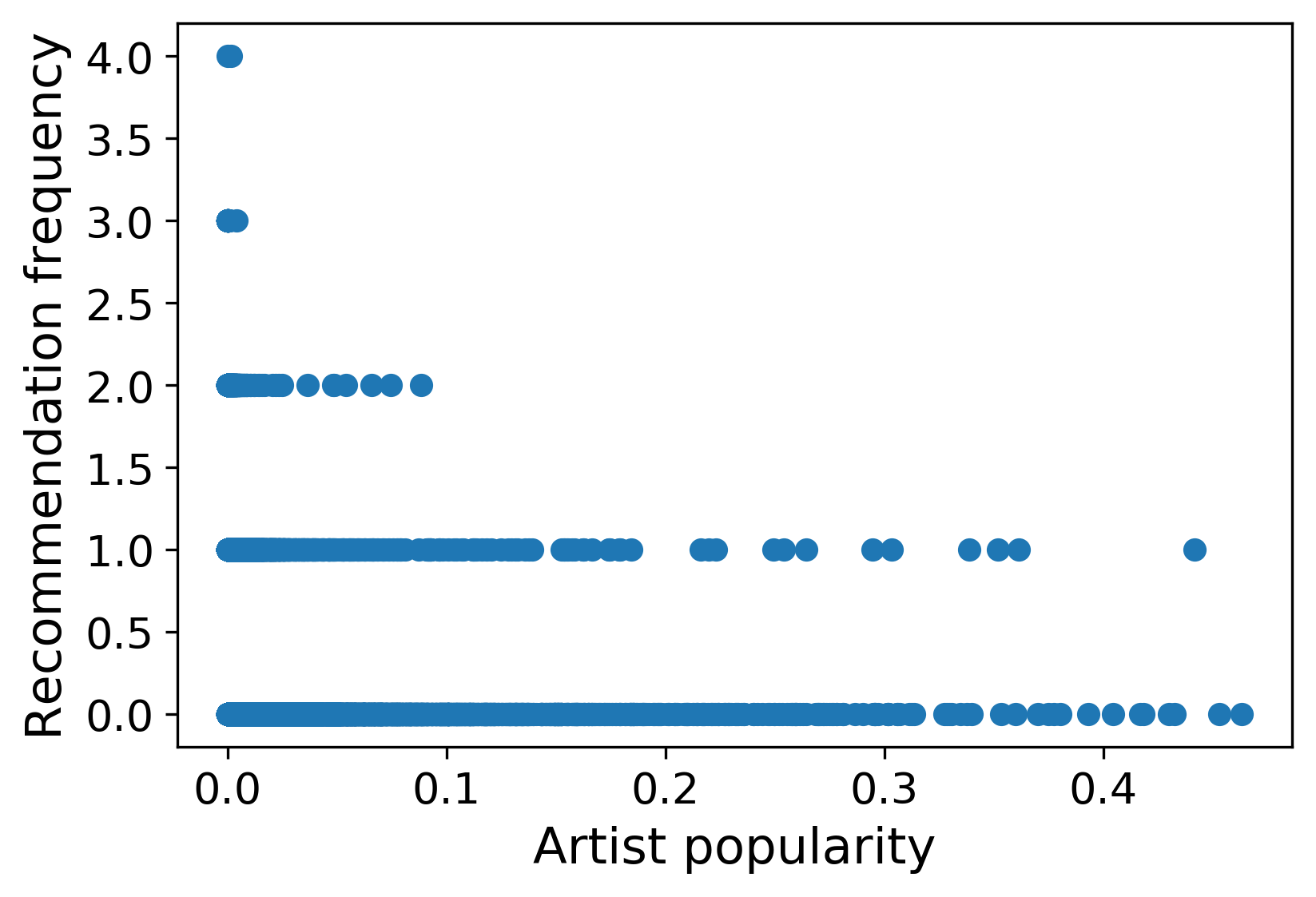}}
~
   \subfloat[MostPopular.]{
      \includegraphics[width=.32\textwidth]{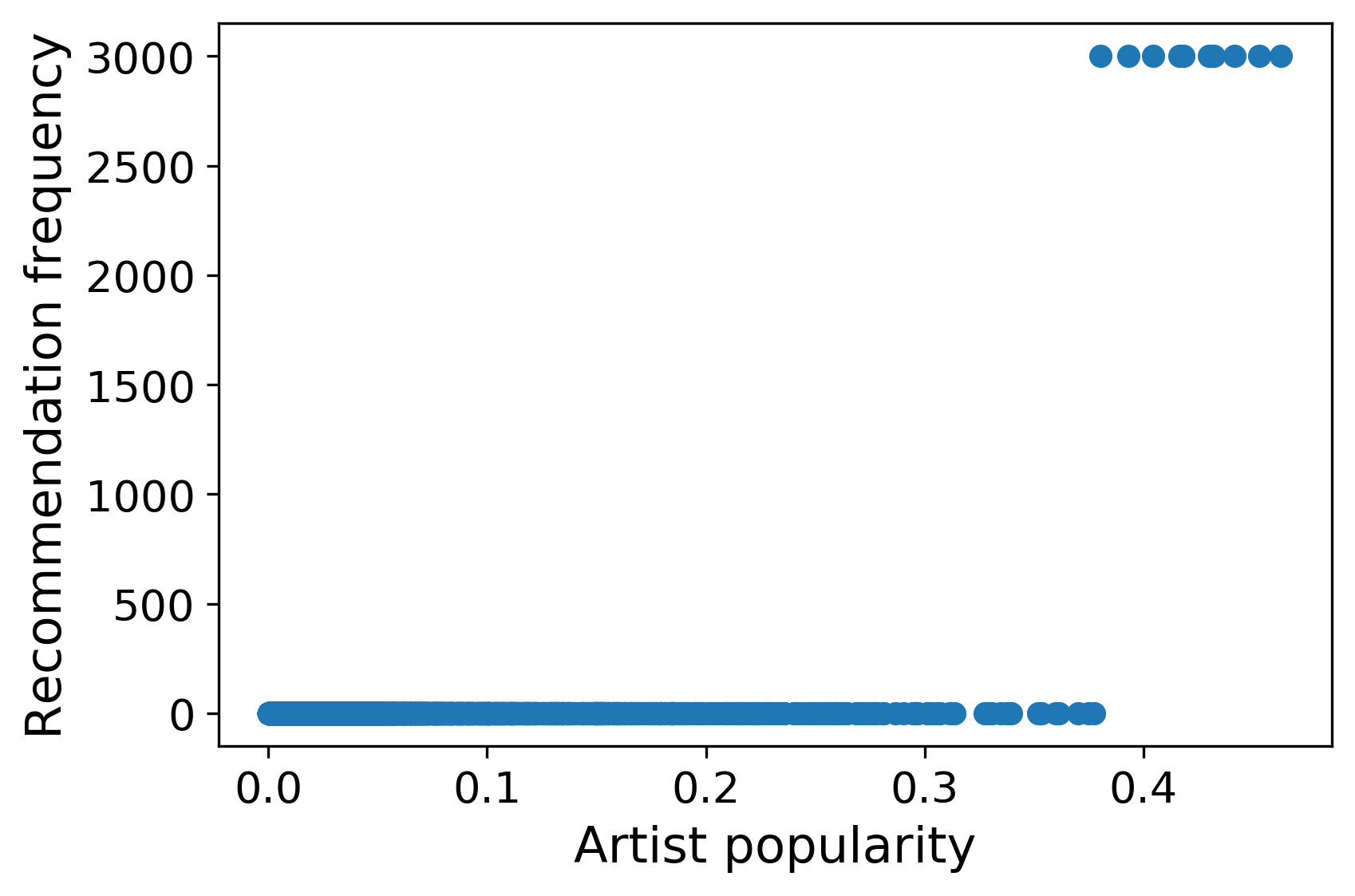}}
~
   \subfloat[UserItemAvg.]{
      \includegraphics[width=.32\textwidth]{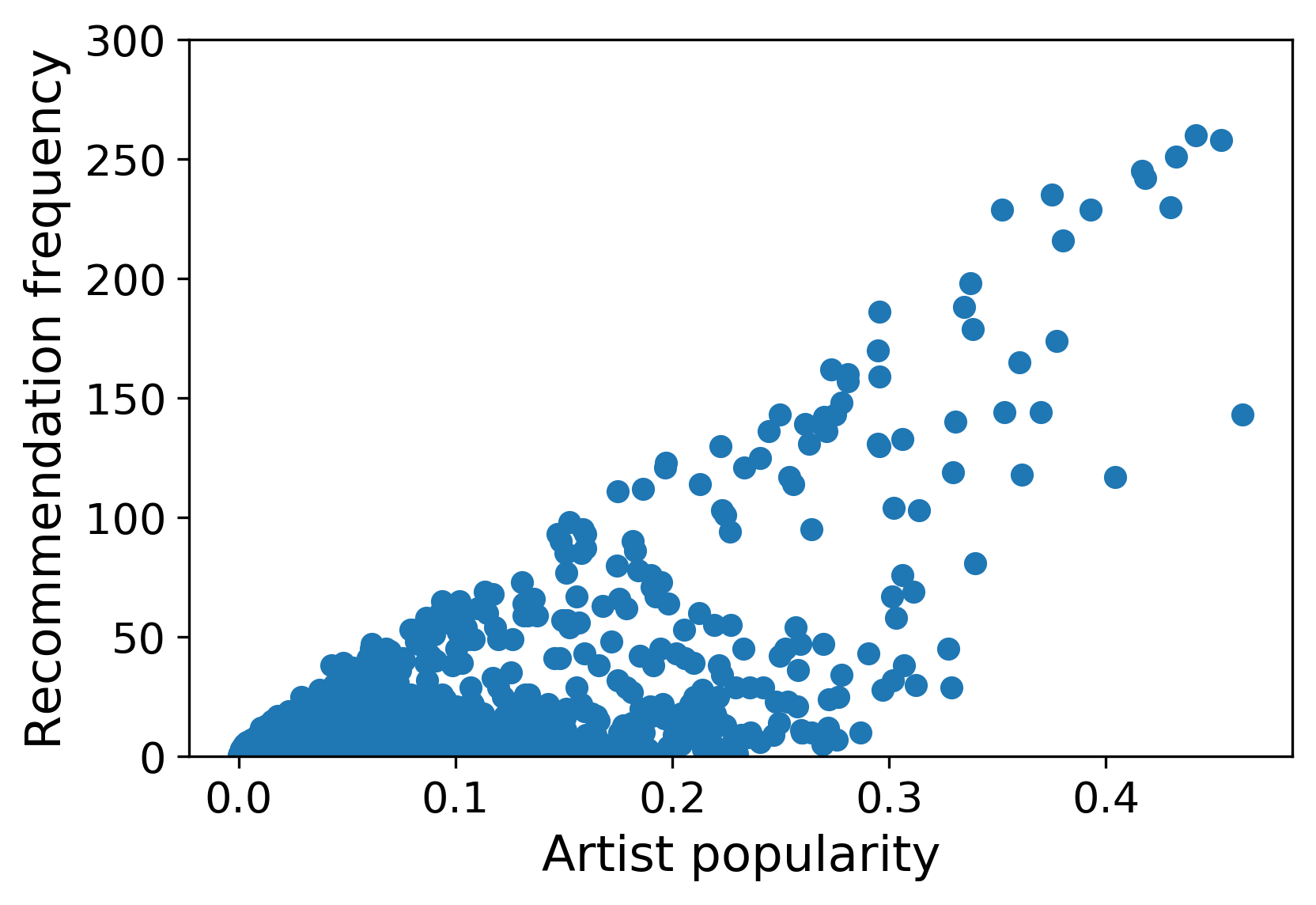}}
\\
   \subfloat[UserKNN.]{
      \includegraphics[width=.32\textwidth]{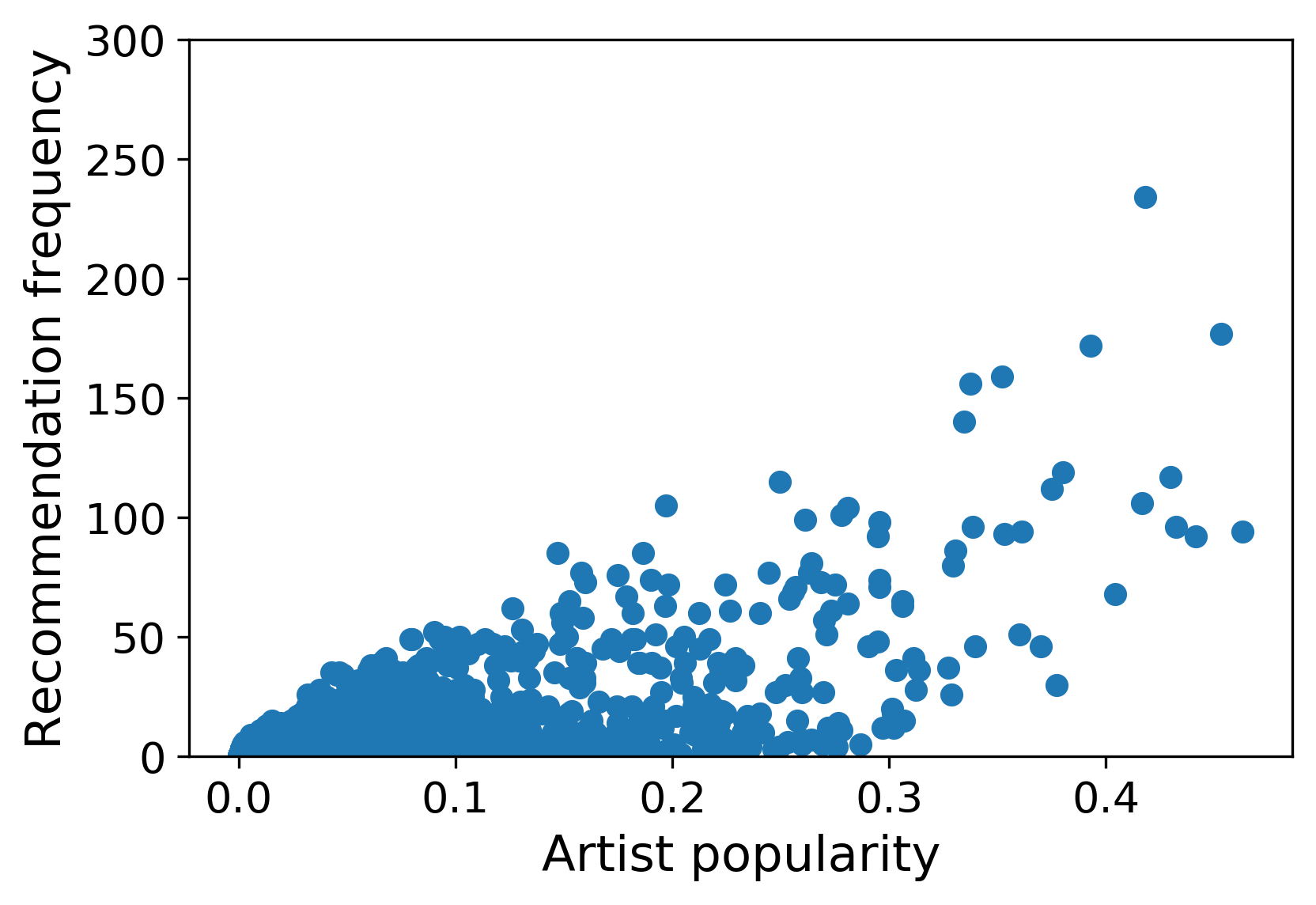}}
~
   \subfloat[UserKNNAvg.]{
      \includegraphics[width=.32\textwidth]{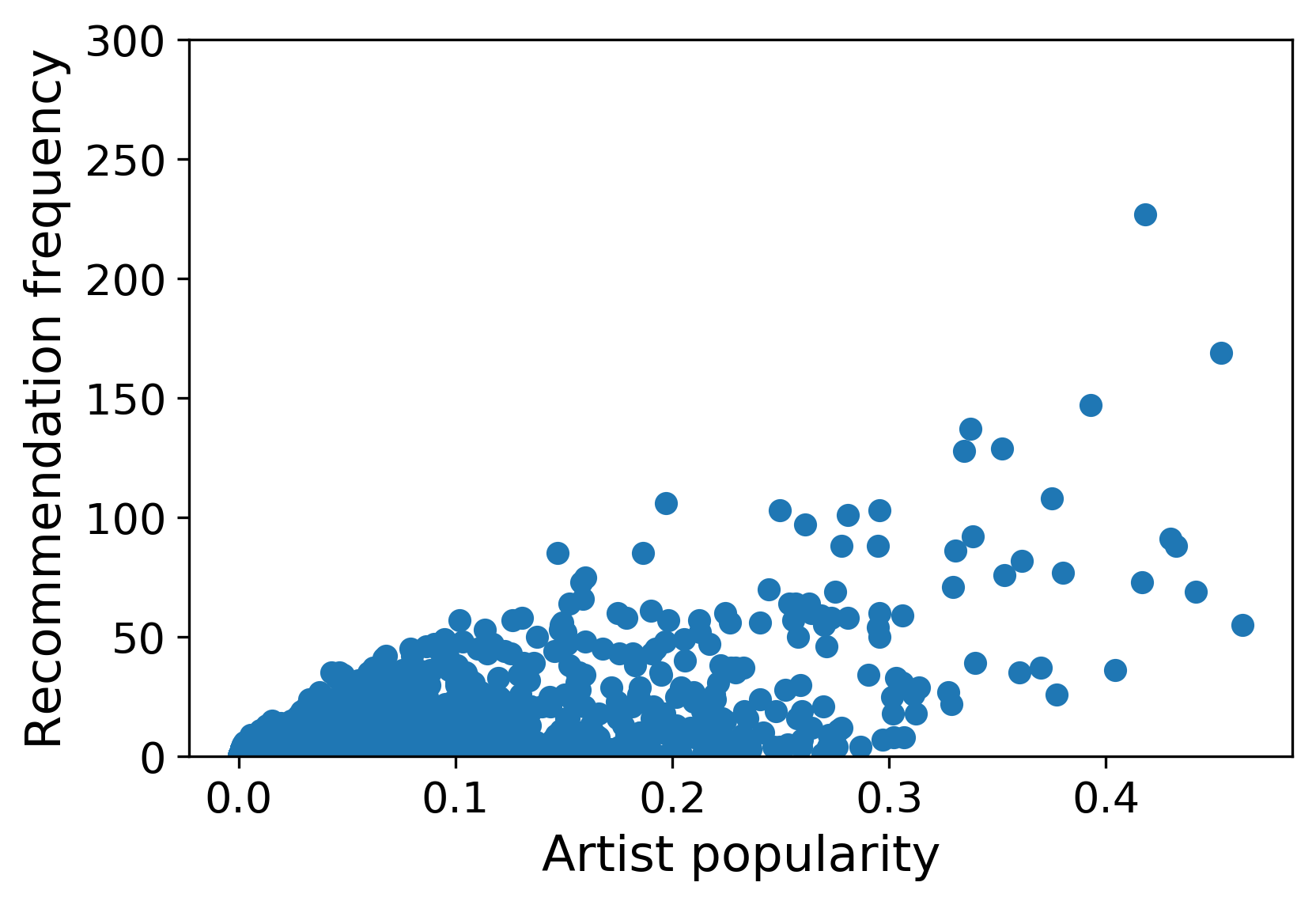}}
~
   \subfloat[NMF.]{
      \includegraphics[width=.32\textwidth]{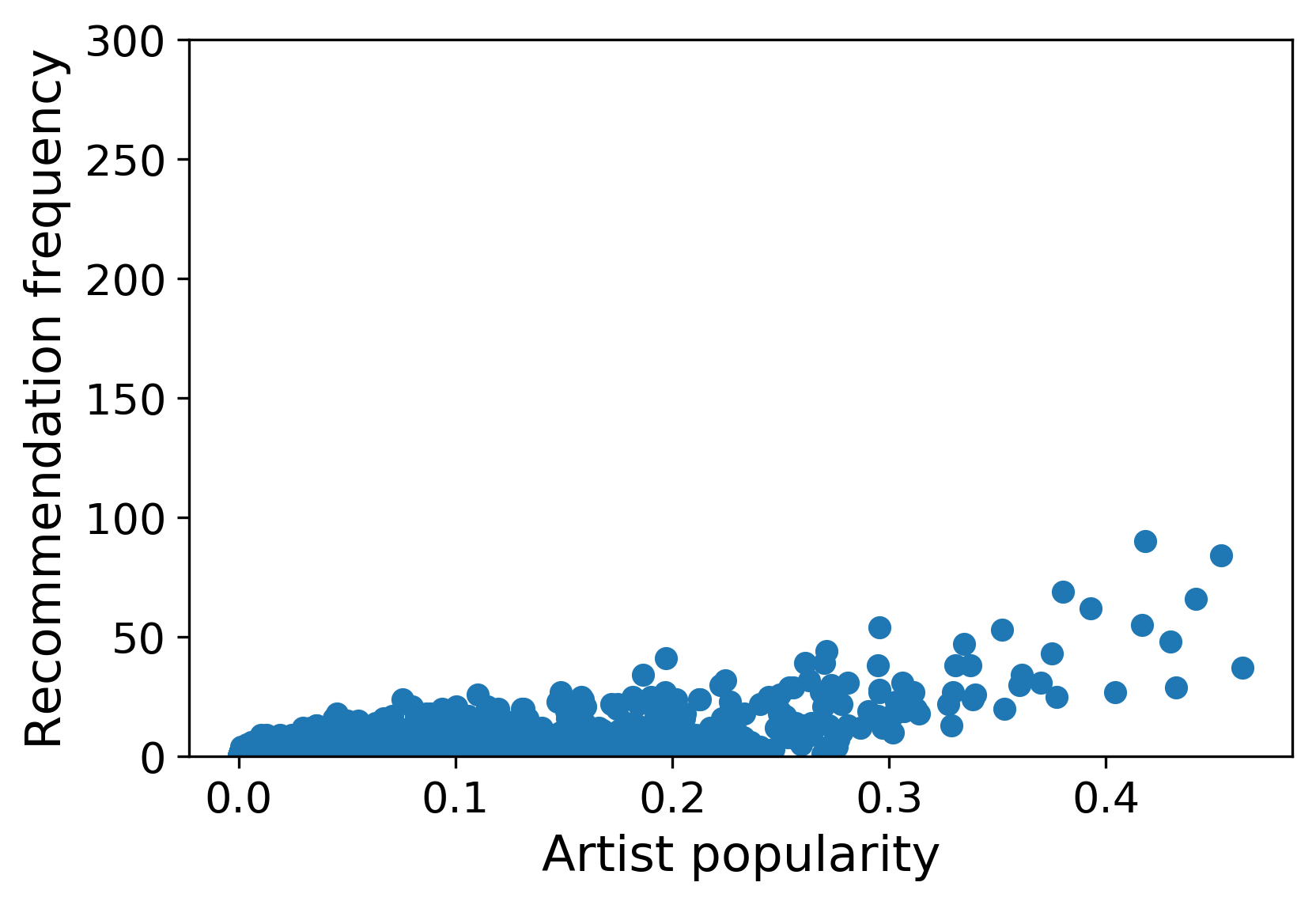}}
   \caption{Correlation of artist popularity and recommendation frequency. For all six algorithms, the recommendation frequency increases with the artist popularity.
   \vspace{-4mm}}
   \label{fig:rec_popularity}
\end{figure}

\section{Popularity Bias in Music Recommendation} \label{sec:rq2}
In this section, we study popularity bias in state-of-the-art music recommendation algorithms. To foster the reproducibility of our study, we calculate and evaluate all recommendations with the Python-based open-source recommendation toolkit Surprise\footnote{\url{http://surpriselib.com/}}. Using Surprise, we formulate our music recommendations as a rating prediction problem, where we predict the preference of a target user $u$ for a target artist $a$. We define the preference of $a$ for $u$ by scaling the listening count of $a$ by $u$ to a range of [0, 1000] as also done in~\cite{schedl2017distance}. We then recommend the top-$10$ artists with the highest predicted preferences.

\para{Recommendation of popular music artists.}
We use the same evaluation protocol (i.e., 80/20 train/test split) and types of algorithms as in~\cite{abdollahpouri2019unfairness}, which includes (i) baseline approaches, (ii) KNN-based approaches, and (iii) Matrix Factorization-based approaches. Specifically, we evaluate three baselines, i.e., Random, MostPopular, and UserItemAvg, which predicts the average listening count in the dataset by also accounting for deviations of $u$ and $a$ (e.g., if $u$ tends to have in general more listening events than the average Last.fm user)~\cite{koren2010factor}. We also evaluate the two KNN-based approaches~\cite{schafer2007collaborative} UserKNN and UserKNNAvg, which is a hybrid combination of UserKNN and UserItemAvg. Finally, we include NMF (Non-Negative Matrix Factorization) into our study~\cite{luo2014efficient}. To reduce the computational effort of our study, in our evaluation, we exclude ItemKNN~\cite{sarwar2001item} as well as SVD++~\cite{sarwar2002incremental} in contrast to~\cite{abdollahpouri2019unfairness}. 
In Figure~\ref{fig:rec_popularity}, we plot the correlation of artist popularity and how often the six algorithms recommend these artists. For all algorithms except for Random, we find a positive correlation, which means that popular items are recommended more often than unpopular items. As expected, this effect is most evident for the MostPopular algorithm and not present at all for the Random algorithm.
It also seems that this popularity bias is not as strong in the case of NMF, which we will investigate further in the next section of this paper.

\begin{figure}[t]
   \centering
   \includegraphics[width=.90\textwidth]{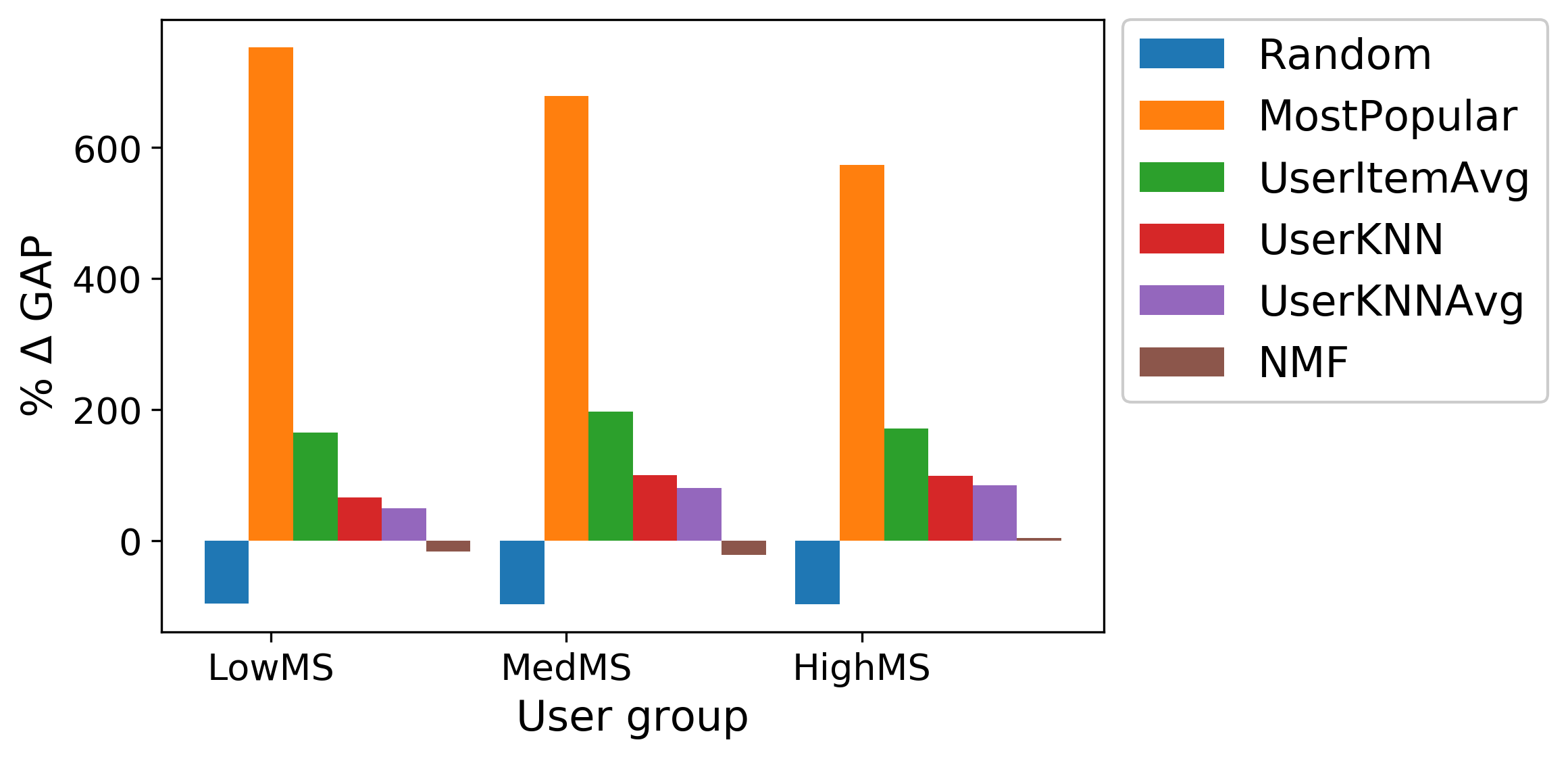}
   \caption{Group Average Popularity ($\Delta$ GAP) of recommendation algorithms for LowMS, MedMS and HighMS. Except for the Random and NMF algorithms, all approaches provide too popular artist recommendations for all three user groups.
   \vspace{-4mm}}
   \label{fig:gap}
\end{figure}

\para{Popularity bias for different user groups.}
To investigate the popularity bias of music recommendations for different user groups (i.e., LowMS, MedMS, and HighMS), we use the Group Average Popularity ($GAP$) metric proposed in~\cite{abdollahpouri2019unfairness}. Here, $GAP(g)_p$ measures the average popularity of the artists in the user profiles $p$ of a specific user group $g$. We also define $GAP(g)_r$, which measures the average popularity of the artists recommended by a recommendation algorithm $r$ to the users of group $g$. For each algorithm and user group, we are interested in the change in $GAP$ (i.e., $\Delta GAP$), which shows how the popularity of the recommended artists differs from the expected popularity of the artists in the user profiles. Hence, $\Delta GAP = 0$ would indicate fair recommendations in terms of item popularity, where fair means that the average artist popularity of the recommendations a user receives matches the average artist popularity in the user's profile. 
It is given by: $\Delta GAP = \frac{GAP(g)_r - GAP(g)_p}{GAP(g)_p}$.

In Figure~\ref{fig:gap}, we plot the $\Delta GAP$ for our six algorithms and three user groups. In contrast to the results presented in~\cite{abdollahpouri2019unfairness}, where the LowMS group (i.e., the niche users) receives the highest values, we do not observe a clear difference between the groups except for MostPopular. We think that this is the case because of the large number of items we have in our Last.fm dataset (i.e., 352,805 artists compared to 3,900 movies in MovieLens). However, in line with Figure~\ref{fig:rec_popularity}, we again find that Random and NMF provide the fairest recommendations.

To further investigate \textit{RQ2}, we analyze the Mean Average Error (MAE)~\cite{willmott2005advantages} results of the four personalized algorithms for our user groups. As shown in Table~\ref{tab:mae}, the LowMS group receives significantly worse (according to a t-test) recommendations than MedMS and HighMS for all algorithms. Interestingly, the MedMS group gets the best recommendations, probably since the users in this group have the largest profiles (i.e., on average 715 artists per user as compared to around 500 for the other two groups).
Across the algorithms, NMF provides the best results. This is especially of interest since NMF also provided the fairest results in terms of artist popularity across the personalized algorithms.

\begin{table}[t]
\setlength{\tabcolsep}{10.0pt}	
\centering
\begin{tabular}{l|llll}
\specialrule{.2em}{.1em}{.1em}
User group & UserItemAvg & UserKNN & UserKNNAvg & NMF    \\\hline
LowMS      & 42.991$^{***}$ & 49.813$^{***}$ & 46.631$^{***}$ & \textbf{38.515$^{***}$} \\
MedMS      & 33.934      & 42.527  & 37.623     & \textbf{30.555} \\
HighMS     & 40.727      & 46.036  & 43.284     & \textbf{37.305} \\\hline
All        & 38.599      & 45.678  & 41.927     & \textbf{34.895} \\
\specialrule{.2em}{.1em}{.1em}
\end{tabular}
\caption{MAE results (the lower, the better) for four personalized recommendation algorithms and our three user groups. The worst (i.e., highest) results are always given for the LowMS user group (statistically significant according to a t-test with $p < .005$ as indicated by $^{***}$). Across the algorithms, the best (i.e., lowest) results are provided by NMF (indicated by bold numbers).
\vspace{-5mm}}
\label{tab:mae}
\end{table}

\section{Conclusion and Future Work}\label{sec:Conclook}
In this paper, we reproduced the study of~\cite{abdollahpouri2019unfairness} on the unfairness of popularity bias in movie recommender systems, which we adopted to the music domain. Similar to the original paper, we find (i) that users only have a limited interest in popular items (\textit{RQ1}) and (ii) that users interested in unpopular items (i.e., LowMS) receive worse recommendations than users interested in popular items (i.e., HighMS). However, we also find that the proposed GAP metric does not provide the same results for Last.fm as it does for MovieLens, probably due to the high number of available items.

For future work, we plan to adapt this GAP metric in order to make it more robust for various domains. Furthermore, we want to study the characteristics of the LowMS users in order to better understand why they receive the worst recommendations and to potentially overcome this with novel algorithms (e.g.,~\cite{ismir_lfm_2019}).

We provide our dataset via Zenodo\footnote{\url{https://doi.org/10.5281/zenodo.3475975}}~\cite{lfm_zenodo} and our source code with all used parameter settings via Github\footnote{\url{https://github.com/domkowald/LFM1b-analyses}}. This work was funded by the Know-Center GmbH (FFG COMET program) and the H2020 projects TRIPLE (GA: 863420) and AI4EU (GA: 825619).

%\bibliographystyle{splncs04}
%\bibliography{bib}

\begin{thebibliography}{10}
\providecommand{\url}[1]{\texttt{#1}}
\providecommand{\urlprefix}{URL }
\providecommand{\doi}[1]{https://doi.org/#1}

\bibitem{abdollahpouri2017controlling}
Abdollahpouri, H., Burke, R., Mobasher, B.: Controlling popularity bias in
  learning-to-rank recommendation. In: Proceedings of the Eleventh ACM
  Conference on Recommender Systems. pp. 42--46. ACM (2017)

\bibitem{abdollahpouri2019unfairness}
Abdollahpouri, H., Mansoury, M., Burke, R., Mobasher, B.: The unfairness of
  popularity bias in recommendation. In: Workshop on Recommendation in
  Multi-stakeholder Environments (RMSE’19), in conjunction with the 13th ACM
  Conference on Recommender Systems, RecSys'19 (2019)

\bibitem{bauer2019global}
Bauer, C., Schedl, M.: Global and country-specific mainstreaminess measures:
  Definitions, analysis, and usage for improving personalized music
  recommendation systems. PloS one  \textbf{14}(6),  e0217389 (2019)

\bibitem{brynjolfsson2006niches}
Brynjolfsson, E., Hu, Y.J., Smith, M.D.: From niches to riches: Anatomy of the
  long tail. Sloan Management Review  \textbf{47}(4),  67--71 (2006)

\bibitem{jannach2015recommenders}
Jannach, D., Lerche, L., Kamehkhosh, I., Jugovac, M.: What recommenders
  recommend: an analysis of recommendation biases and possible countermeasures.
  User Modeling and User-Adapted Interaction  \textbf{25}(5),  427--491 (2015)

\bibitem{koren2010factor}
Koren, Y.: Factor in the neighbors: Scalable and accurate collaborative
  filtering. ACM Transactions on Knowledge Discovery from Data (TKDD)
  \textbf{4}(1), ~1 (2010)

\bibitem{ismir_lfm_2019}
Kowald, D., Lex, E., Schedl, M.: Modeling artist preferences for personalized
  music recommendations. In: {Proceedings of the Late-Breaking-Results Track of
  the 20th Annual Conference of the International Society for Music Information
  Retrieval}. ISMIR '19 (2019)

\bibitem{lfm_zenodo}
Kowald, D., Schedl, M., Lex, E.: {LFM User Groups} (2019).
  \doi{10.5281/zenodo.3475975}

\bibitem{luo2014efficient}
Luo, X., Zhou, M., Xia, Y., Zhu, Q.: An efficient non-negative
  matrix-factorization-based approach to collaborative filtering for
  recommender systems. IEEE Transactions on Industrial Informatics
  \textbf{10}(2),  1273--1284 (2014)

\bibitem{ricci2011introduction}
Ricci, F., Rokach, L., Shapira, B.: Introduction to recommender systems
  handbook. In: {Recommender Systems Handbook}, pp. 1--35. Springer (2011)

\bibitem{sarwar2002incremental}
Sarwar, B., Karypis, G., Konstan, J., Riedl, J.: {Incremental Singular Value
  Decomposition Algorithms for Highly Scalable Recommender Systems}. In:
  Proceedings of the Fifth International Conference on Computer and Information
  Science. vol.~27, p.~28 (2002)

\bibitem{sarwar2001item}
Sarwar, B.M., Karypis, G., Konstan, J.A., Riedl, J., et~al.: Item-based
  collaborative filtering recommendation algorithms. {WWW}  \textbf{1},
  285--295 (2001)

\bibitem{schafer2007collaborative}
Schafer, J.B., Frankowski, D., Herlocker, J., Sen, S.: Collaborative filtering
  recommender systems. In: {The Adaptive Web}, pp. 291--324. Springer (2007)

\bibitem{schedl2016lfm}
Schedl, M.: {The LFM-1B Dataset for Music Retrieval and Recommendation}. In:
  Proceedings of the 2016 ACM on International Conference on Multimedia
  Retrieval. pp. 103--110. ICMR '16, ACM, New York, NY, USA (2016)

\bibitem{schedl2017distance}
Schedl, M., Bauer, C.: Distance-and rank-based music mainstreaminess
  measurement. In: Adjunct Publication of the 25th Conference on User Modeling,
  Adaptation and Personalization. pp. 364--367. ACM (2017)

\bibitem{schedl2018current}
Schedl, M., Zamani, H., Chen, C.W., Deldjoo, Y., Elahi, M.: Current challenges
  and visions in music recommender systems research. International Journal of
  Multimedia Information Retrieval  \textbf{7}(2),  95--116 (2018)

\bibitem{willmott2005advantages}
Willmott, C.J., Matsuura, K.: {Advantages of the mean absolute error (MAE) over
  the root mean square error (RMSE) in assessing average model performance}.
  Climate Research  \textbf{30}(1),  79--82 (2005)

\end{thebibliography}

\end{document}